\documentclass[conference]{IEEEtran}
\IEEEoverridecommandlockouts

\usepackage{cite}
\usepackage{amsmath,amssymb,amsfonts}
\usepackage{algorithmic}
\usepackage{graphicx}
\usepackage{textcomp}
\usepackage{xcolor}
\usepackage{booktabs}
\usepackage{xurl}
\usepackage{amssymb}
\usepackage{caption}
\usepackage{graphicx} 
\def\BibTeX{{\rm B\kern-.05em{\sc i\kern-.025em b}\kern-.08em
    T\kern-.1667em\lower.7ex\hbox{E}\kern-.125emX}}
\begin{document}

\title{Cross-Field Interface-Aware Neural Operators for Multiphase Flow Simulation\\
\thanks{\textsuperscript{*}Equal contribution, \textsuperscript{\dag}Corresponding author}
}

\author{
\IEEEauthorblockN{Zhenzhong Wang\textsuperscript{*}, Xin Zhang\textsuperscript{*}, Jun Liao, Min Jiang\textsuperscript{\dag}}
\IEEEauthorblockA{\textit{
    School of Informatics, Xiamen University
}\\
Email: minjiang@xmu.edu.cn}
}
\maketitle

\begin{abstract}
Multiphase flow simulation is critical in science and engineering but incurs high computational costs due to complex field discontinuities and the need for high-resolution numerical meshes. While Neural Operators (NOs) offer an efficient alternative for solving Partial Differential Equations (PDEs), they struggle with two core challenges unique to multiphase systems: spectral bias caused by spatial heterogeneity at phase interfaces, and the persistent scarcity of expensive, high-resolution field data. This work introduces the Interface Information Aware Neural Operator (IANO), a novel architecture that mitigates these issues by leveraging readily obtainable interface data (e.g., topology and position). Interface data inherently contains the high-frequency features not only necessary to complement the physical field data, but also help with spectral bias. IANO incorporates an interface-aware function encoding mechanism to capture dynamic coupling, and a geometry-aware positional encoding method to enhance spatial fidelity for pointwise super-resolution. Empirical results across multiple multiphase flow cases demonstrate that IANO achieves significant accuracy improvements (up to $\sim$10\%) over existing NO baselines. Furthermore, IANO exhibits superior generalization capabilities in low-data and noisy settings, confirming its utility for practical, data-efficient $\text{AI}$-based multiphase flow simulations.
\end{abstract}

\section{Introduction}
Multiphase flow, defined as the simultaneous flow of two or more distinct phases (e.g., liquid-gas and liquid-liquid) that coexist and interact within a system, plays a pivotal role across a wide spectrum of fields, including the chemical industry, environmental protection, and life sciences~\cite{brennen2005fundamentals,cao2025gas,zhao2024single,dong2025multi}. Due to the complex dynamics, field discontinuities, and interphase interactions, using numerical simulation to accurately compute the multiple physics fields (e.g., velocity and temperature) of multiphase flows necessitates the use of extremely fine meshes to approximate the solutions of the governing Partial Differential Equations (PDEs). Consequently, the computational cost becomes prohibitively high~\cite{brennen2005fundamentals,prosperetti2009computational}. 

\begin{figure}[t]
	\centering
\includegraphics[width=0.908\linewidth]{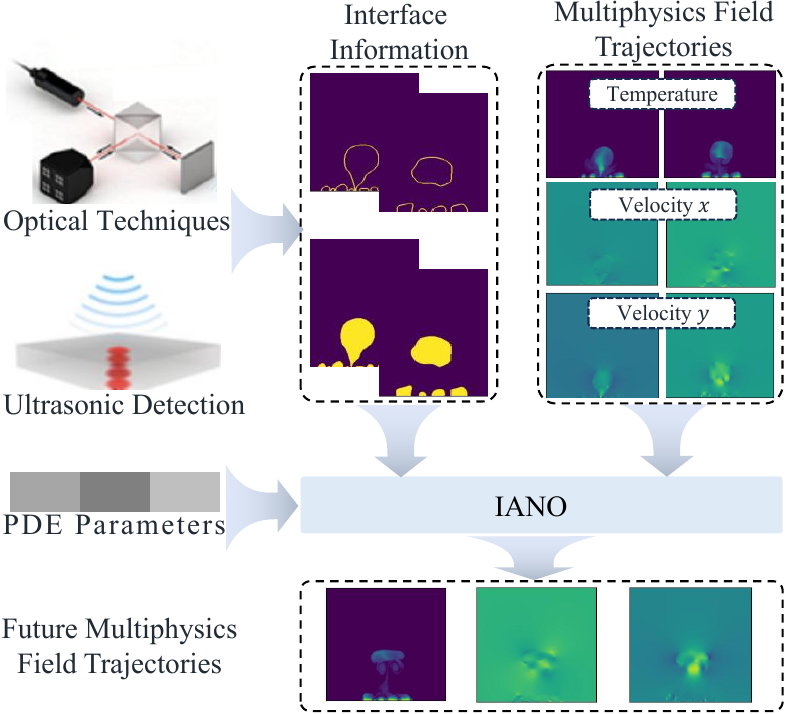}
\vspace{-0.37cm}
	\caption{IANO combines interface data as auxiliary variables with the multiphysics field data for multiphase flow simulation.}
	\label{fig:intro}
    \vspace{-0.59cm}
\end{figure}

Over the past few years, many deep learning techniques have been proposed to provide efficient alternatives~\cite{karniadakis2021physics,azizzadenesheli2024neural,gong2024adversarial,yang2025pegnetphysicsembeddedgraphnetwork,Cao_Liu_Wang_Xu_Ye_Tan_Jiang_2024,cao2025interpretable}. Among others, neural operators learn the infinite-dimensional mapping from functional parameters to the solution~\cite{kovachki2023neural,cao2024laplace,raonic2023convolutional,wei2023super}, which requires training only once and predicts solutions through a single forward computation, significantly accelerating the process of solving PDEs. For instance, DeepONet~\cite{lu2019deeponet} uses a branch network and a trunk network to process input functions and query coordinates. This architecture has been proven to approximate any nonlinear operators with a sufficiently large network. MIONet~\cite{mionet} extends DeepONets to solve problems with multiple input functions. Fourier neural operator (FNO)~\cite{lifourier} operates in the spectral space using the Fast Fourier Transform, which achieves a good cost-accuracy trade-off.

Despite the remarkable progress, multiphase flow simulation exhibits unique challenges. Firstly, due to the inherent differences in physical properties of phases and substances, multiphase flow systems exhibit \emph{spatial heterogeneity} --- i.e., physical quantities (such as velocity, temperature, etc.) in multiphase flow systems are unevenly distributed and discontinuous. This exacerbates the \emph{spectral bias} of neural operators, i.e., they struggle to capture high-frequency features, such as drastic local physical quantity changes. In particular, the drastic physical quantities are often at the phase interface of multiphase flow, while they are of the greatest interest, as they directly reflect the system characteristics, e.g., the heat transfer behavior in thermal management systems~\cite{lee2009low,ordonez2024thermal}. Secondly, accurately representing these key physical features at the interface demands massive high-resolution data. Unfortunately, the scarcity of these expensive data constrains the predictive capabilities of neural operators.

This work introduces the Interface Information Aware Neural Operator (IANO), a novel neural operator specifically designed for multiphase flow simulations. Particularly, we leverage interface data to mitigate the challenges posed by spectral bias and the scarcity of data. As depicted in Fig. \ref{fig:intro}, techniques such as optical techniques, ultrasonic detection, etc, can directly capture crucial interface information~\cite{mao2020physics,poletaev2020bubble,murai2010ultrasonic,buhendwa2021inferring,murai2001three}, including its topology, position, and trajectory. 
Compared to field quantities that are difficult to measure directly, this interface information is more readily obtainable. Numerical methods can also calculate the phase field to capture the dynamics of the interface. This interface data not only serves as a complement to the physical field data but also intuitively reflects the complex interactions between different phases. They directly characterize the location and geometry of interfaces inherently containing high-frequency features, contributing to mitigating spectral bias.


Specifically, IANO has two main components. We first propose a mechanism that processes multiple physical fields alongside the interface information, capturing not only the interactions among various physical variables but also their dynamic coupling with the interface data. Second, we craft a geometry-aware positional encoding method to establish relationships between spatial positions, interface features, physical variables, and PDE parameters. By meticulously computing these pointwise relationships, this mechanism enhances spatial fidelity.
Our contributions are summarized as follows:

\begin{enumerate}
    \item IANO mitigates the spectral bias and data scarcity issues by leveraging interface data as a crucial complement. Integrating interface data obtained from experimental techniques or numerical simulations could pave an effective way for accurate multiphase flow simulations.
    \item Two novel mechanisms are elaborated. First, an interface-aware multiple functions encoding mechanism captures the coupling between physical variables and interfaces, enabling IANO to capture high-frequency physical features at the interface.
    Second, a geometry-aware positional encoding method establishes the relationships between interface features, physical variables, and spatial positions, thereby achieving pointwise super-resolution prediction.
    \item  Experimental results across multiple cases demonstrate significant improvements over existing baselines, achieving up to a $\sim$10.0\% increase in accuracy. Furthermore, IANO showcased superior generalization capabilities in low-data super-resolution and noisy data settings.
\end{enumerate}

\section{Related Work}

Neural operators have emerged as a transformative paradigm for approximating mappings between infinite-dimensional function spaces for solving PDEs. They can be broadly categorized into the following three types: 1) Spectral-Domain Operators: The FNO \cite{lifourier} pioneered learning in the spectral domain through Fourier transforms, enabling efficient parameter-to-solution mappings in frequency space. Subsequent extensions, such as Geo-FNO \cite{li2023fourier}, address complex geometries by incorporating coordinate transformations, while domain decomposition variants \cite{hu2023augmented, grady2022towards} facilitate multi-scale resolution. Although highly computationally efficient, these spectral methods face inherent limitations when dealing with discontinuous fields and heterogeneous input modalities. 2) Branch-Trunk Architectures: DeepONet \cite{lu2019deeponet} established a foundational framework that decouples input processing via a branch network from query point encoding via a trunk network, offering universal approximation guarantees. Physics-informed variants \cite{wang2021learning, wang2022improved} enhance generalization capabilities by integrating PDE-constrained optimization, and MIONet \cite{mionet} extends this architecture to handle systems with multiple input functions. While demonstrating notable flexibility across various PDEs, these methods typically encounter scaling challenges when applied to high-dimensional parameter spaces. 3) Transformer-Based Operators: Recent advancements have leveraged transformer architectures to achieve enhanced geometric adaptability in neural operators.
OFormer addresses multi-scale challenges through structured attention patterns~\cite{li2023transformer}. Galerkin Transformer \cite{cao2021choose}, for instance, reinterprets attention mechanisms as learnable integral kernels, establishing theoretical connections to traditional operator learning. More recently, CoDA-NO \cite{coda-no} introduced a breakthrough by treating physical variables as infinite-dimensional tokens, enabling native handling of function space inputs.

Despite the progress, they often face challenges when applied to multiphase flow. A primary concern is the spectral bias --- they struggle to capture high-frequency features in multiphase flows, such as sharp interfaces and drastic local changes. Furthermore, neural operators are inherently data-hungry. This impedes the predictive performance of neural operators in multiphase flow, as acquiring high-fidelity training data, particularly at interfaces, is computationally prohibitive.



\begin{figure*}[t]
	\centering	\includegraphics[width=0.878\linewidth]{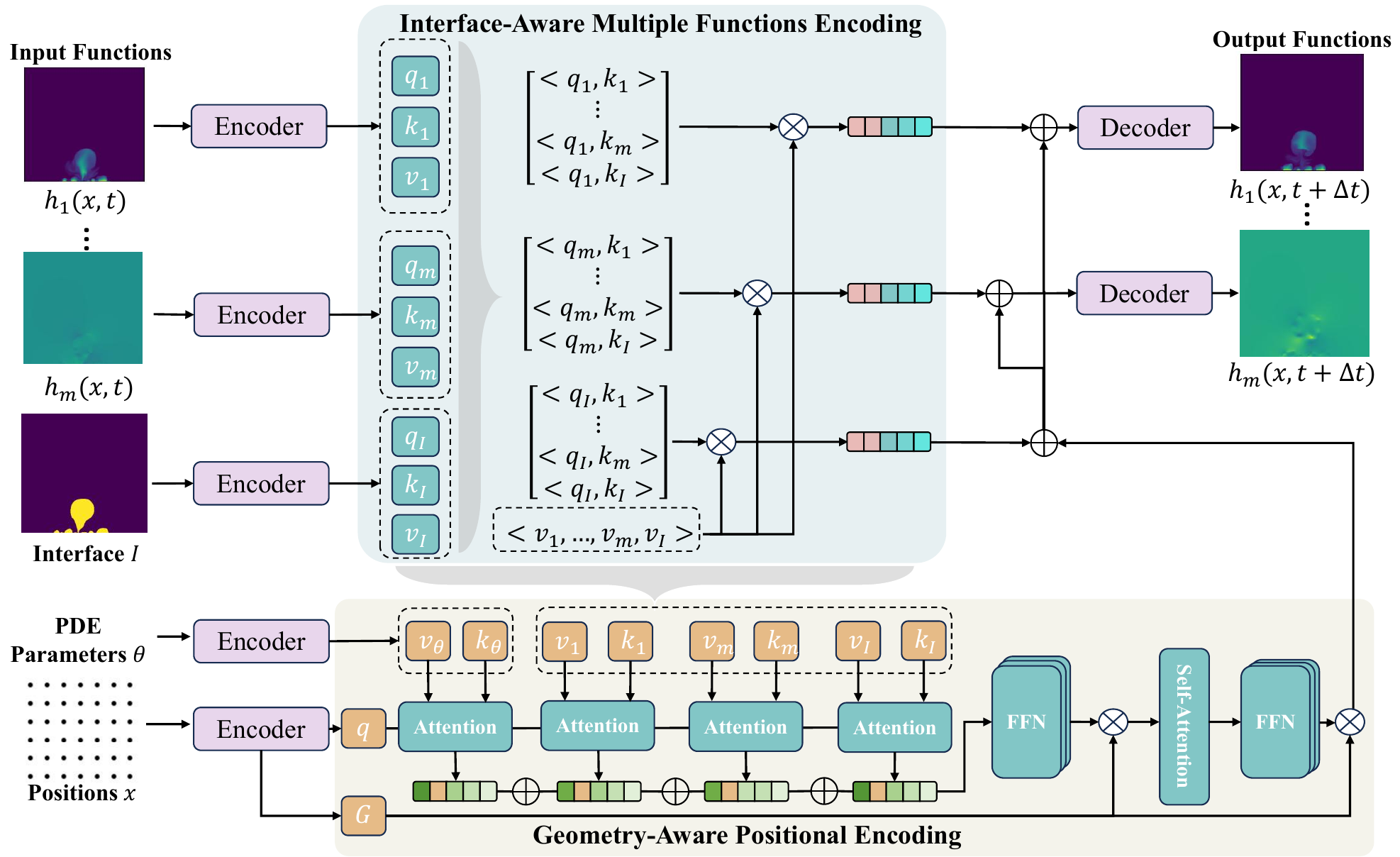}
	\caption{IANO's architecture: 1) The interface-aware multiple functions encoding mechanism jointly encodes both multiphysics fields and interface information to generate an interface-aware function embedding for each field. 2) The geometry-aware positional encoding mechanism produces positional embedding by explicitly linking multiphysics fields and interfaces to their positions. IANO integrates the function embeddings with the positional embeddings for multiphase flow prediction.
    }
    \vspace{-0.25cm}
	\label{fig:pipeline}
\end{figure*}

\section{Methodology}
This section formalizes the multiphase flow simulation problem and provides a detailed description of our proposed IANO's architecture and its key components.

\subsection{Problem Formulation}

We consider multiphase flow systems governed by PDEs with parameters $\theta$  within a spatial domain $\Omega \subset \mathbb{R}^d$ over a temporal interval $t \in [0, T]$. Let the input function space be denoted by $\mathcal{A} = \mathcal{H}_1 \times \cdots \times \mathcal{H}_m \times \theta \times \mathcal{I}$. Here, each $\mathcal{H}_j$ ($j = 1, \ldots, m$) represents a distinct physical field (e.g., velocity, pressure, or source terms defined over $\Omega$). The term $\theta \in \mathbb{R}^p$ denotes system parameters (e.g., viscosity ratios), and $I \in \mathcal{I}$ represents the interface geometry.

Our primary objective is to learn a neural operator $\mathcal{G}: \mathcal{A} \rightarrow \mathcal{U}$ that maps an input tuple $a = (h_1(\cdot,t), \ldots, h_m(\cdot,t), \theta, I(t))$ to the time-dependent solution field $u(\cdot, t+\Delta t) \in \mathcal{U}$. The input tuple $a$ comprises physical fields $h_i\in \mathcal{H}_i$, system parameters $\theta$, and interface information $I$. The output solution field $u: \Omega \times [0, T] \rightarrow \mathbb{R}^n$ is required to satisfy the governing PDEs along with interface-dependent constraints. The operator $\mathcal{G}$ explicitly incorporates both the input physical fields and the interface information to accurately predict the evolution of multiphase flow dynamics.

\subsection{Framework Overview}

Here, we present an overview of our proposed IANO, as illustrated in Fig. \ref{fig:pipeline}. Multiphase flow simulation inherently involves the coupling of multiple physical fields and exhibits significant spatial heterogeneity. The intricate interplay between interface evolution and field dynamics creates strongly nonlinear interdependencies. To effectively capture this complex synergy, we introduce two main components: an interface-aware multiple functions encoding mechanism and a geometry-aware positional encoding mechanism.

Firstly, the interface-aware multiple functions encoding explicitly models both inter-field interactions and their dynamic coupling with interface geometries. Its goal is to generate an interface-aware function embedding for each input physical field. This enables the simultaneous learning of cross-field interactions and localized physics-interface interactions, which are modulated by interface motion. 

Secondly, we propose a geometry-aware positional encoding mechanism. This mechanism establishes a precise positional context by explicitly linking physical variables to their interface positions and input function dependencies, thereby enabling the accurate resolution of multiscale dynamics. The learned geometry-aware positional embeddings, in conjunction with the interface-aware function embeddings, are then used to predict the evolution of multiphase flow. 
In the following parts, we provide the details of the two main components.

\subsection{Interface-Aware Multiple Functions Encoding}
\label{sec:interface_aware_functions_encoding}

Multiphase flow typically involves complex coupling effects between various input functions (e.g., velocity and temperature fields) and the intricate interaction between flow characteristics and interface dynamics (e.g., the influence of interface tension on the velocity field). Our proposed interface-aware multiple functions encoding scheme is designed to capture these two effects within a unified manner.

First, the input functions and the interface information are mapped into a high-dimensional vector space. To preserve the functional nature of the heterogeneous input data, we utilize FNO \cite{lifourier} and U-Net~\cite{takamoto2022pdebench} to create the representations for input functions and interface information, respectively. Specifically, we denote these transformations as $\mathbf{H}_i = f_{\text{FNO}}(h_i)$ for input function $h_i$ and $\mathbf{H}_I = f_{\text{U-Net}}(I)$ for interface information $I$. Each representation then undergoes $L_2$ normalization to eliminate the influence of dimensional units, yielding normalized representations $\frac{\mathbf{H}_i}{||\mathbf{H}_i||_2}$ and $\frac{\mathbf{H}_I}{||\mathbf{H}_I||_2}$.

Subsequently, we employ a cross-attention mechanism to model the intricate coupling between multiple physical input functions and interface dynamics in multiphase flow. Given $M$ normalized physical field representations $\{\mathbf{H}_1, \mathbf{H}_2, \ldots, \mathbf{H}_M\}$ and a normalized interface representation $\mathbf{H}_I$, we transform them into query-key-value triplets. For each representation, we generate corresponding queries ($q$), keys ($k$), and values ($v$) within a latent dimension $d$:
\begin{align}
\mathbf{H}_i &\rightarrow (q_i, k_i, v_i) \in \mathbb{R}^{d}, \quad \text{for } i=1, \dots, M, \\
\mathbf{H}_I &\rightarrow (q_I, k_I, v_I) \in \mathbb{R}^{d}.
\end{align}
We then compute an inner product matrix to quantify the pairwise interactions between all physical fields and the interface dynamics. For the $i$-th physical field, the dot product values measuring its synergy with other field variables and the interface are calculated as: $[ \langle q_i, k_1 \rangle, \ldots, \langle q_i, k_M \rangle, \langle q_i, k_I \rangle ]^\top$. Here, $\langle q_i, k_i \rangle$ captures the spatial self-correlation of the $i$-th field, $\langle q_i, k_j \rangle$ (for $i \neq j$) models the coupling effects between the target field and other physical fields, and $\langle q_i, k_I \rangle$ quantifies the interaction between the field and interface dynamics. The dot product $\langle q_i, k_j \rangle$ is realized by an integral: $\int k_{\psi}(q,k)k \mathrm{d}k$, where the integral can be efficiently discretized using quadrature rules.

Based on this dot product matrix, we use the cross-attention mechanism to update features for the $i$-th physical field. The resulting interface-aware function embedding for $\mathbf{H}_i$ is computed as:
\begin{equation}
\begin{split}
\mathbf{H}'_i = & \sigma\left( \left[ \frac{\langle q_i, k_1 \rangle}{||q_i||_2 ||k_1||_2}, \ldots, \frac{\langle q_i, k_M \rangle}{||q_i||_2 ||k_M||_2}, \frac{\langle q_i, k_I \rangle}{||q_i||_2 ||k_I||_2} \right]^\top \right) \\&\cdot \left[ \frac{v_1}{||v_1||_2}, \ldots, \frac{v_M}{||v_M||_2}, \frac{v_I}{||v_I||_2} \right]^\top / \sqrt{d_{\text{latent}}},
\end{split}
\end{equation}
where $d_{\text{latent}}$ is a scaling factor and $\sigma(\cdot)$ is the softmax function. This cross-attention mechanism aggregates information from all input functions and the additional interface information. As shown in Fig. \ref{fig:pipeline}, we also incorporate a skip connection through an identity mapping to ensure that essential information is preserved. This module introduces interface information to force the IANO to learn the dependence between multiple physical fields and interfaces, thus alleviating spectral bias.

\subsection{Geometry-Aware Positional Encoding}
\label{sec:interface_aware_positional_encoding}

This part details our geometry-aware positional encoding mechanism, which explicitly links physical variables and interface to their respective positions. 

In addition to the query-key-value triplets of the input functions and interface information, we encode the PDE parameters $\theta$ into a similar query-key-value triplet using an MLP: $\mathbf{H}_P = f_{\text{MLP}}(\theta)$. This parameter representation is also normalized, $\mathbf{H}_P = \mathbf{H}_P / ||\mathbf{H}_P||_2$, and transformed into $(q_P, k_P, v_P) \in \mathbb{R}^d$.

We then compute the attention output with the spatial positional information $q_x$ as the query. The initial attention formulation is:
\begin{equation}
    \mathbf{Z}_x = \sum_{i \in \mathcal{K}} \frac{q_x \cdot k_i}{\sum_{l \in \mathcal{K}} q_x \cdot k_l} \cdot v_i,
    \label{eq:attention}
\end{equation}
where $\mathcal{K} = \{k_1, \dots, k_M, k_I, k_P\}$ represents the set of all keys (from $M$ physical fields, interface, and parameters). To improve computational efficiency, we introduce a normalization factor $\alpha_t = \left(\sum_{l \in \mathcal{K}} q_x \cdot k_l\right)^{-1}$, allowing the attention to be reformulated as:
\begin{equation}
    \mathbf{Z}_x = \sum_{i \in \mathcal{K}} \alpha_t(q_x \cdot k_i) \cdot v_i = \alpha_t q_x \cdot \left(\sum_{i \in \mathcal{K}} k_i \otimes v_i\right).
    \label{eq:efficient_attention}
\end{equation}
Given that we have $M+2$ sequences of key-value pairs (from $M$ physical fields, interface information, and PDE parameters), the cross-attention is formulated as:
\begin{equation}\label{eq:cross_attention2}
    \mathbf{Z}_x = q_x + \frac{1}{M+2} \sum_{l=1}^{M+2} \alpha^l_t q_x \cdot \left(\sum_{i_l=1}^{N_l} k_{i_l} \otimes v_{i_l}\right),
\end{equation}
where $\alpha^l_t = \frac{1}{\sum_{s=1}^{N_l} q_x \cdot k_s}$ is the normalization coefficient for each embedded sequence $l$. This cross-attention mechanism effectively harmonizes multiple sources of information, ensuring that the geometry-aware positional encoding is well-integrated with the physical context.

Subsequently, a self-attention layer is applied to the enriched query features, further refining the representation:
\begin{equation}
    \mathbf{Z}'_x = \sum_{i } \alpha_t(q_x \cdot k_i) \cdot v_i
    \label{eq:self_attention}
\end{equation}
where $q_x$, $k_x$, and $v_x$ are computed from the embedding $\mathbf{Z}_x$ as follows:
\begin{equation}
    q_x = W_q \mathbf{Z}_x, \quad k_x = W_k \mathbf{Z}_x, \quad v_x = W_v \mathbf{Z}_x.
\end{equation}
Here, $W_q, W_k, W_v$ are learnable weight matrices. In summary, the components explicitly link physical variables and interfaces to their positions, thereby enabling the accurate resolution of multiscale dynamics.


\subsection{Learning Objective}

The learning objective of IANO is to predict physical field quantities at the next time step $t+\Delta t$ by leveraging the learned embedding $\mathbf{H}^{'}$ and $\mathbf{Z}^{'}$, which encapsulates the core information to describe the target physical fields. Specifically, the FNO serves as the decoder for the prediction. The FNO decoder is tasked with mapping the latent embedding $\mathbf{H}^{'}\oplus\mathbf{Z}^{'}$ to the target physical fields $h_1^*, \ldots, h_M^*$. The training is accomplished by minimizing the discrepancy between the predicted physical fields and their corresponding ground truth fields. The loss function $\mathcal{L}$ for this minimization is the Mean Squared Error (MSE), which quantifies the square root of the average squared difference between predicted and true values across all spatial points and all predicted fields. 
\begin{equation}
    \mathcal{L} = \sqrt{\frac{1}{M} \sum_{i=1}^{M} \text{MSE}(\hat{h}_{i}, h_{i}^*)},
\end{equation}
where $\hat{h}_{i}$ represents the $i$-th predicted physical field, $h_{i}^*$ is the $i$-th ground truth physical field.


\section{Experiments}


\subsection{Evaluation Protocol} We conduct experiments on five representative cases of multiphase flow, including a single bubble, saturated pool boiling, subcooled pool boiling, flow boiling gravity, and 3D pool boiling International Space Station (ISS) gravity. The interplay between bubble dynamics and heat transfer makes the simulation extremely challenging. The five cases are generated based on the following equations~\cite{bubbleml}, 
\paragraph*{Momentum Transport Equation}
\begin{equation}
    \frac{\partial\vec{u}}{\partial t}+\vec{u}\cdot\nabla\vec{u}=-\frac{1}{\rho'}\nabla P+\nabla\cdot\left[\frac{\mu'}{\rho'}\frac{1}{\text{Re}}\nabla\vec{u}\right]+\frac{\vec{g}}{\text{Fr}^{2}}+\vec{S}_{u}^{\Gamma}+S_{P}^{\Gamma}.
    \label{eq:momentum}
\end{equation}

\paragraph*{Energy Transport Equation}
\begin{equation}
    \frac{\partial T}{\partial t}+\vec{u}\cdot\nabla T=\nabla\cdot\left[\frac{\alpha'}{\text{Re}~\text{Pr}}\nabla T\right]+S_{T}^{\Gamma}.
    \label{eq:energy}
\end{equation}

\paragraph*{Continuity Equation}
\begin{equation}
    \nabla \cdot \vec{u} = -\dot{m} \nabla(\rho^{\prime})^{-1} \cdot \vec{n},
\end{equation}
where $\vec{u}$ is the velocity vector field, $P$ is the pressure, and $T$ is the temperature throughout the domain. The terms $\rho^{\prime}$, $\mu^{\prime}$, and $\alpha^{\prime}$ in the equations are local fluid properties that have been non-dimensionalized. Scaled fluid properties, such as $\rho^{\prime}$, represent the local value of the phase scaled by the corresponding value in the liquid. \text{Re}, \text{Pr}, and \text{Fr} are Reynolds number, Prandtl number, and Froude number, respectively. $\vec{g}$ and $\dot{m}$ represent the gravity vector and the mass transfer rate at the interface. $\vec{n}$ is the unit normal vector to the liquid-vapor interface.
Furthermore, at the liquid-vapor interface $\Gamma$, the source terms $\vec{S}_{u}^{\Gamma}$ and $S_{T}^{\Gamma}$ model the effects of evaporation on velocity and temperature, while $S_{P}^{\Gamma}$ accounts for the pressure jump caused by surface tension.




The five cases used were generated by numerically solving the above non-dimensionalized transport equations.  The interface data is simulated using Flash-X through the level set method~\cite{dubey2022flash}. The detailed description of each case can be found in Appendix B.


\begin{figure*}[t]
\centering
\includegraphics[width=0.9\linewidth]{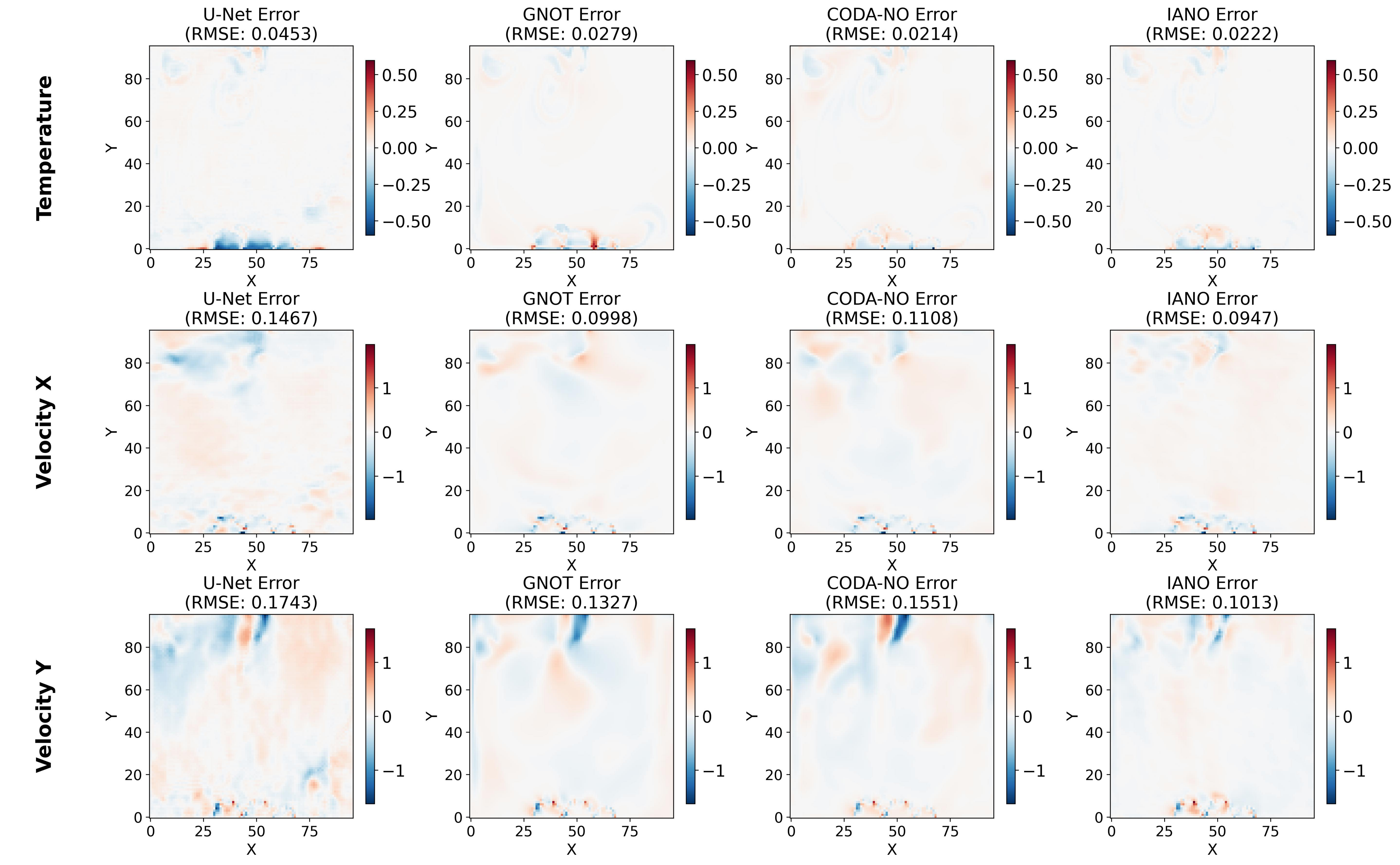}
\vspace{-0.24cm}
	\caption{The RMSE error plot for Subcooled Pool Boiling at 2× resolution. 
    IANO consistently achieves the lowest RMSE.
    \vspace{0.4cm} 
    }
    \vspace{-0.54cm}
	\label{fig:errorvis}
\end{figure*}

\begin{table*}[h]
\centering
\caption{The main results of predicted multiple fields on several multiphase flow cases. CODA-NO exhibits unstable training behavior on the Single Bubble case, resulting in failure to produce valid results. }
\resizebox{0.8\linewidth}{!}
{
    \begin{tabular}{@{}llcccccccccc@{}}
    \toprule
    \multicolumn{2}{c}{Model} & \multicolumn{2}{c}{MIONet} & \multicolumn{2}{c}{U-Net}       & \multicolumn{2}{c}{CODA-NO} & \multicolumn{2}{c}{GNOT} & \multicolumn{2}{c}{IANO}        \\
    Cases& Fields  & RMSE         & IRMSE       & RMSE           & IRMSE          & RMSE         & IRMSE        & RMSE        & IRMSE      & RMSE           & IRMSE          \\ \midrule
    Subcooled       & $T$     & 0.046        & 0.160       & 0.035          & 0.129          & 0.036        & 0.134        & 0.037       & 0.139      & \textbf{0.030} & \textbf{0.118} \\
                    & $u_x$   & 0.225        & 0.670       & 0.183          & 0.647          & 0.197        & 0.674        & 0.208       & 0.644      & \textbf{0.173} & \textbf{0.629} \\
                    & $u_y$   & 0.216        & 0.521       & 0.162          & 0.479          & 0.175        & 0.501        & 0.213       & 0.491      & \textbf{0.146} & \textbf{0.459} \\
    Gravity         & $T$     & 0.119        & 0.182       & \textbf{0.059} & \textbf{0.102} & 0.105        & 0.167        & 0.119       & 0.181      & 0.067          & 0.110          \\
                    & $u_x$   & 0.247        & 0.670       & \textbf{0.215} & \textbf{0.359} & 0.237        & 0.393        & 0.238       & 0.383      & 0.282          & 0.430          \\
                    & $u_y$   & 0.174        & 0.521       & \textbf{0.150} & \textbf{0.252} & 0.170        & 0.269        & 0.174       & 0.270      & 0.170          & 0.271          \\
    Saturated       & $T$     & 0.052        & 0.114       & 0.045          & 0.106          & 0.047        & 0.109        & 0.050       & 0.113      & \textbf{0.042} & \textbf{0.101} \\
                    & $u_x$   & 0.512        & 0.709       & 0.403          & 0.644          & 0.442        & 0.676        & 0.470       & 0.679      & \textbf{0.361} & \textbf{0.607} \\
                    & $u_y$   & 0.543        & 0.720       & 0.404          & 0.611          & 0.439        & 0.662        & 0.486       & 0.671      & \textbf{0.358} & \textbf{0.552} \\
    Single Bubble   & $T$     & 0.027        & 0.048       & 0.015          & 0.040          & —            & —            & 0.009       & 0.031      & \textbf{0.006} & \textbf{0.021} \\
                    & $u_x$   & 0.100        & 0.206       & 0.060          & 0.150          & —            & —            & 0.067       & 0.154      & \textbf{0.059} & \textbf{0.133} \\
                    & $u_y$   & 0.096        & 0.185       & 0.065          & 0.150          & —            & —            & 0.068       & 0.142      & \textbf{0.063} & \textbf{0.133} \\
    3D ISS  Gravity& $T$     & 0.033& 0.056& 0.033& 0.063& 0.016& 0.036& 0.047& 0.103& \textbf{0.014}& \textbf{0.032}\\
                    & $u_x$   & 0.113& 0.246& 0.103& 0.219& 0.096& 0.225& 0.125& 0.259& \textbf{0.090}& \textbf{0.212}\\
                    & $u_y$   & 0.089& 0.188& 0.075& 0.172& 0.079& 0.181& 0.105& 0.214& \textbf{0.070}& \textbf{0.165}\\
                    & $u_z$   & 0.107& 0.229& 0.086& 0.202& 0.091& 0.210& 0.120& 0.242& \textbf{0.083}& \textbf{0.196}\\ \bottomrule
    \end{tabular}
}
    \vspace{-0.15cm}
\label{tab:mainresults}
\end{table*}

\paragraph{Baselines} We compare our IANO against several recently proposed state-of-the-art baselines, detailed as follows:
\begin{itemize}
\item \textbf{MIONet}~\cite{mionet}: This architecture extends the traditional DeepONet by incorporating multiple branch networks to handle multiple input functions, enabling its application to multiphysics fields.
\item \textbf{U-Net}~\cite{takamoto2022pdebench}: A variant of the well-known U-Net architecture, this model specifically integrates batch normalization layers to enhance training stability and convergence for solving multiphysics fields.
\item \textbf{GNOT}~\cite{hao2023gnot}: A transformer-based neural operator to address the irregular meshes, multiple input functions, and multi-scale scenarios.
\item \textbf{CODA-NO}~\cite{coda-no}: A pre-trained neural operator specifically developed for handling multiphysics systems.
\end{itemize}

\paragraph{Evaluation Metrics.} We utilize two metrics: the Root Mean Square Error (RMSE) and the Interface Root Mean Square Error (IRMSE). RMSE measures the predicted physical field's accuracy across the entire computational domain. 
 $   \mathrm{RMSE} = \sqrt{
    \frac{1}{\lvert \Omega \rvert}
    \sum_{x \in \Omega}
    \bigl(\hat{h}(x) - h^*(x)\bigr)^2
}$
, here $\Omega$ represents all positional points within the domain, while $\hat{h}$ and $h^*$ correspond to the predicted and true field values, respectively.

IRMSE specifically quantifies the prediction error along the interfaces. It not only serves as a localized accuracy assessment at boundaries but also offers insights into the model's ability to capture high-frequency details. $
\mathrm{IRMSE} = \sqrt{
    \frac{1}{\lvert \Gamma \cap \Omega \rvert}
    \sum_{x \in \Gamma \cap \Omega}
    \bigl(\hat{h}(x) - h^*(x)\bigr)^2
}$,
where $\Gamma \cap \Omega$ denotes the subset of positional points from the interface that fall within the computational domain $\Omega$.

We predict the temperature and velocity fields for the five cases mentioned above. It is important to note that in numerical simulations, pressure primarily serves to correct velocity fields rather than acting as a truly accurate physical quantity itself. Since pressure is computed by solving a Poisson equation, and Poisson solvers can inherently lack robustness, obtaining highly precise pressure values is often challenging. Therefore, we do not predict pressure. Following previous works~\cite{bonneville2025accelerating,li2024local,li2023phase}, all compared baselines perform prediction in a step-wise manner.
The parameter settings and the training details are given in Appendix A.

\subsection{Results and Discussion}

We evaluated the IANO's prediction capabilities against baselines across five multiphase flow scenarios, assessing performance with RMSE and IRMSE for temperature ($T$) and velocity fields ($u_x$, $u_y$). As shown in Table \ref{tab:mainresults}, IANO consistently delivered significant performance advantages across most scenarios. For example, in subcooled pool boiling, IANO's temperature field predictions achieved an RMSE of 0.030 and an IRMSE of 0.118, marking improvements of roughly 14.3\% and 8.5\% over the leading baseline, U-Net. IANO's most decisive advantage emerged in the single bubble case, where its temperature field yielded an RMSE of just 0.006 and an IRMSE of 0.021, representing substantial improvements of approximately 33.3\% and 32.3\% over the GNOT model. While U-Net showed a slight edge in the gravity case, IANO generally achieved the lowest errors or remained highly competitive across all state-of-the-art methods. More importantly, the significant improvement in IRMSE highlights its capabilities in capturing interface regions' high-frequency features in multiphase flow simulation. We also visualized the prediction error as shown in Fig. \ref{fig:errorvis} and Appendix G.

In addition, we conducted the prediction in a rollout manner, i.e., given an initial state $u(t_{k-1})$, we compute the next state $u(t_k)$ using the operator. We then continue this process iteratively to obtain a sequence of predicted states. From the results in Appendix C, our IANO method consistently achieves the best performance across most fields. Regarding the computational efficiency (see Appendix G), IANO can strike a commendable balance between its superior predictive performance and reasonable computational efficiency.


\begin{figure*}[ht]
  \centering
  \begin{minipage}[t]{0.24\linewidth}
    \centering    \includegraphics[width=\linewidth]{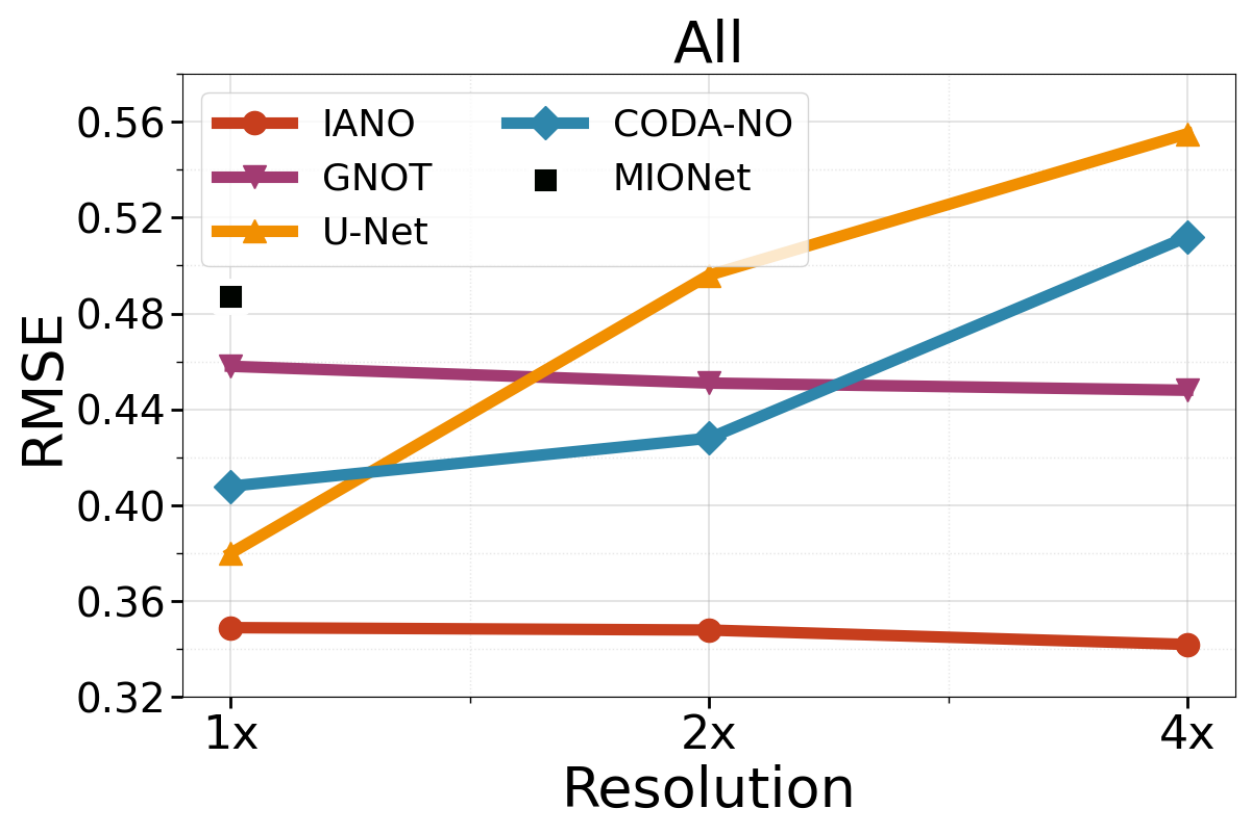}
    \vspace{-0.7cm}
    \caption*{(a)}
  \end{minipage}
  \hfill
  \begin{minipage}[t]{0.24\linewidth}
    \centering
    \includegraphics[width=\linewidth]{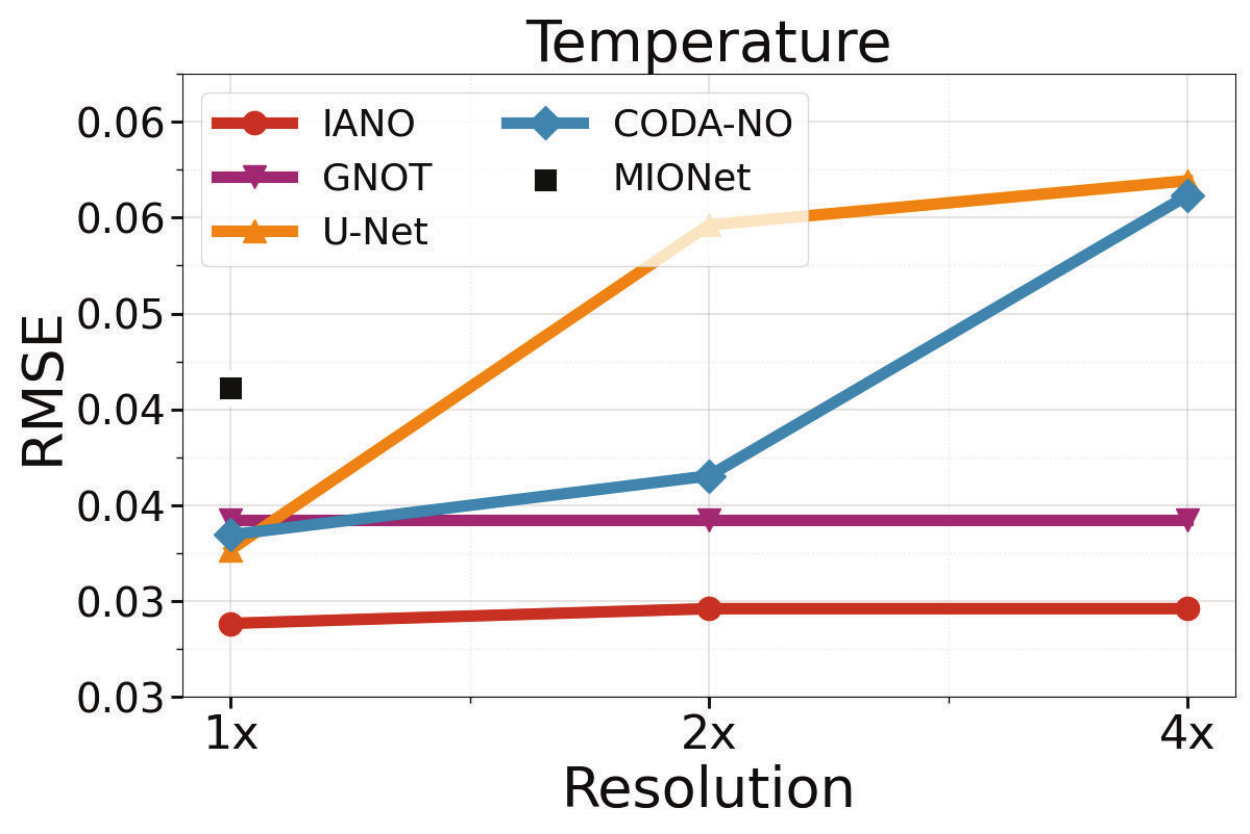}
       \vspace{-0.7cm}
    \caption*{(b)}
  \end{minipage}
  \hfill
  \begin{minipage}[t]{0.24\linewidth}
    \centering
    \includegraphics[width=\linewidth]{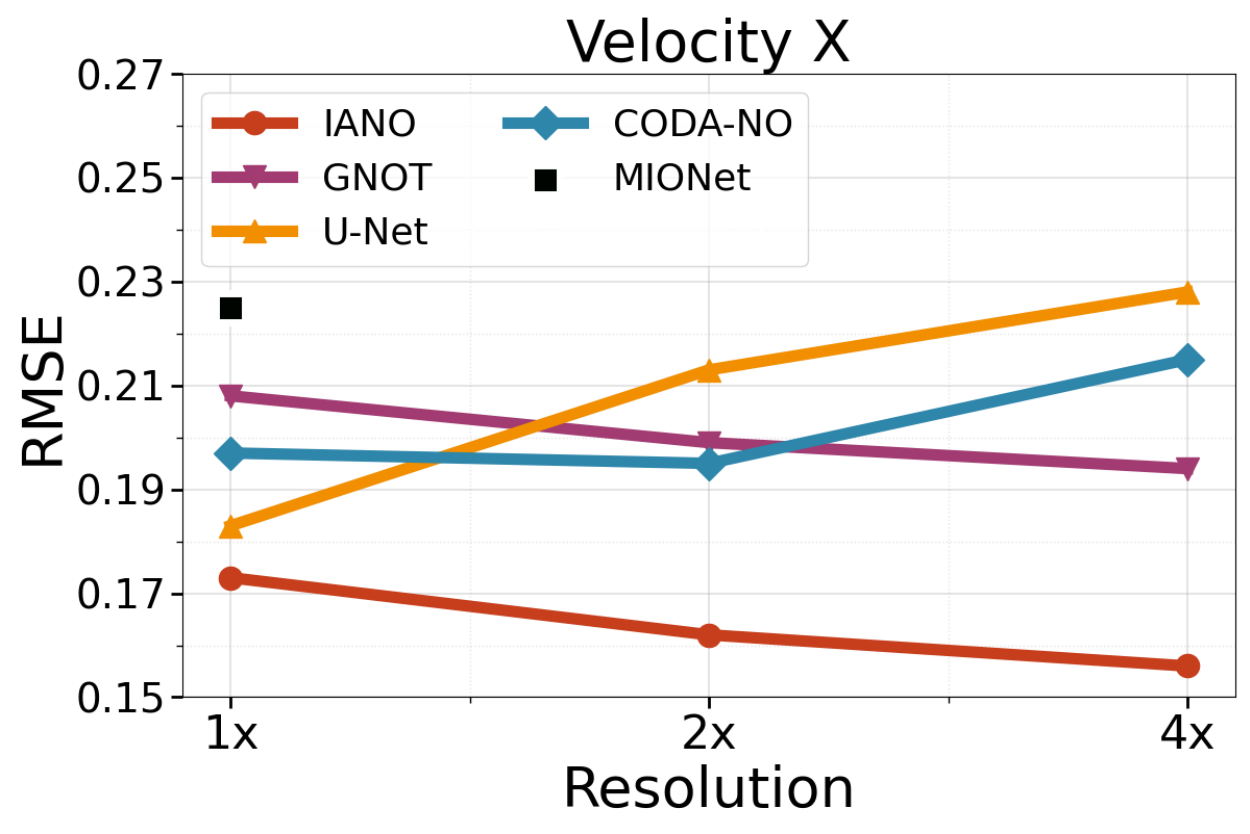}
        \vspace{-0.7cm}
    \caption*{(c)}
  \end{minipage}
  \hfill
  \begin{minipage}[t]{0.24\linewidth}
    \centering
    \includegraphics[width=\linewidth]{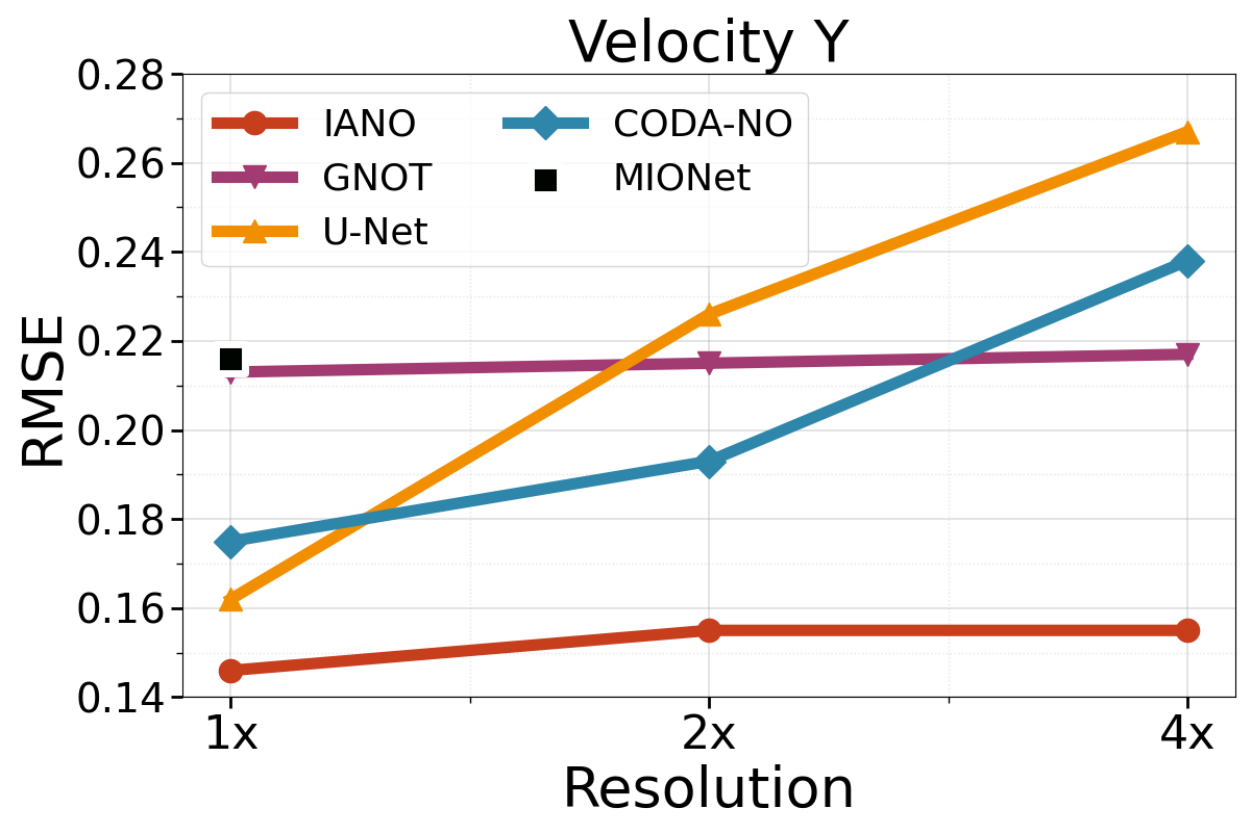}
        \vspace{-0.7cm}
    \caption*{(d)}
  \end{minipage}
  \vspace{-0.25cm}
  \caption{IANO's super-resolution performance. (a) Average RMSE across all fields. (b)-(d) RMSE for temperature, velocity in the x-direction, and velocity in the y-direction, respectively. MIONet, whose branch networks rely on fixed sampling locations, does not support super‑resolution; hence, only one data point is shown.  }
    \vspace{-0.45cm}
  \label{fig:SR}
\end{figure*}

\subsection{Super-Resolution Performance}

This section examines the performance of our proposed IANO method for super-resolution physical field prediction with 2x and 4x resolutions. Fig. \ref{fig:SR} (a) -- Fig. \ref{fig:SR} (d) depicts the RMSE along with different resolutions, and the detailed experimental results are presented in Appendix D. In the results, 1x signifies that the training and test data resolutions are identical, while 2x and 4x indicate that the test data resolution is two and four times, respectively, that of the training data.
The results show IANO maintains leading prediction accuracy across all cases. In the 4x super-resolution task for the temperature field, IANO's RMSE remained exceptionally low at 0.031, with its IRMSE only slightly increasing to 0.132. In contrast, U-Net's RMSE significantly rose to 0.060, and CODA-NO's to 0.059. For the GNOT model, with an RMSE of 0.037 and IRMSE of 0.163 at 4x, fell short of IANO. For the velocity fields, IANO likewise recorded the lowest RMSE values at 4x super-resolution, demonstrating a clear performance gap against models like U-Net and CODA-NO at the same resolutions.

\begin{figure}[t]
  \centering
  \begin{minipage}[t]{0.49\linewidth}
    \centering
    \includegraphics[width=\linewidth]{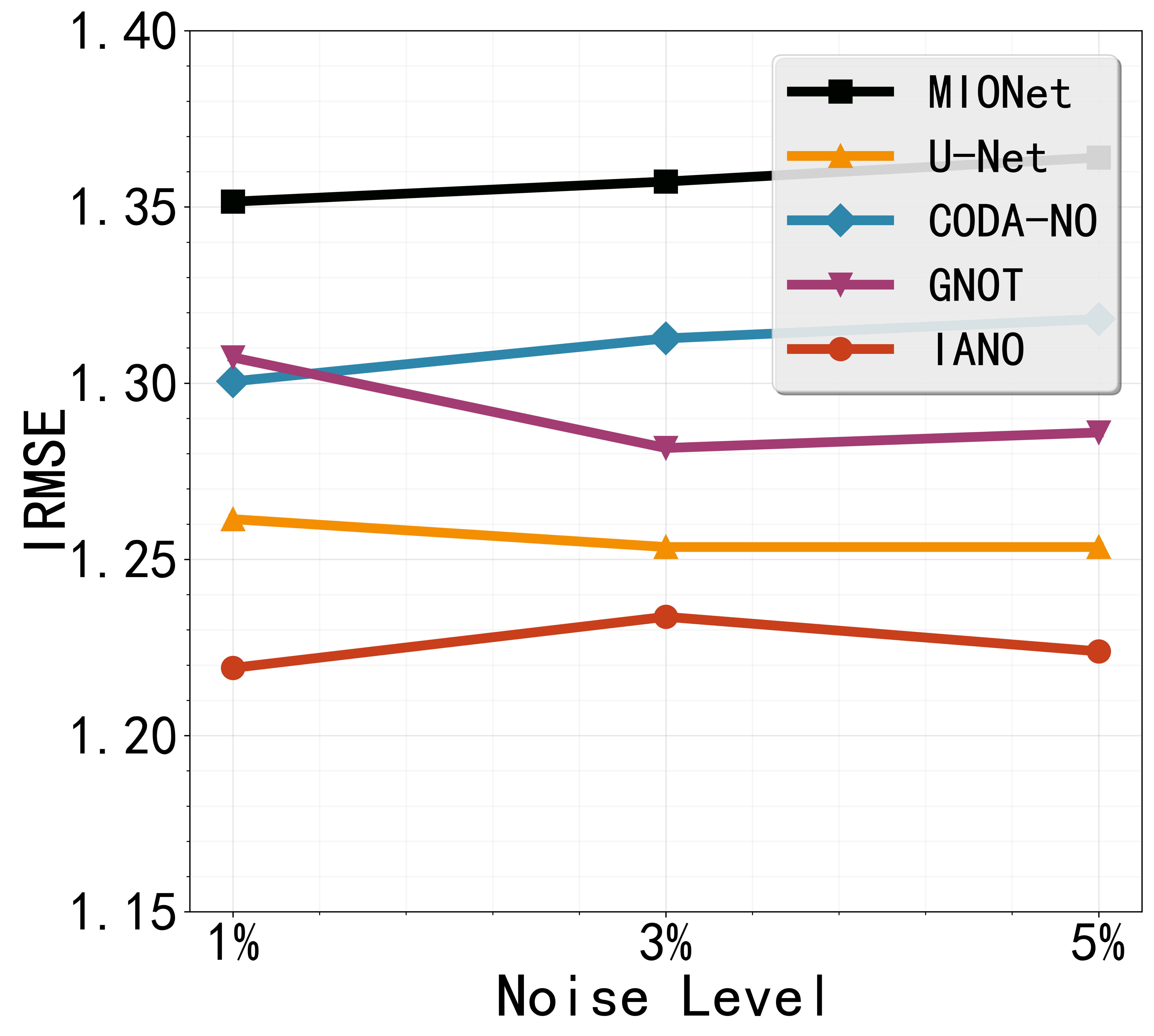}
         \vspace{-0.7cm}
    \caption*{(a)}
  \end{minipage}
  \hfill
  \begin{minipage}[t]{0.49\linewidth}
    \centering
    \includegraphics[width=\linewidth]{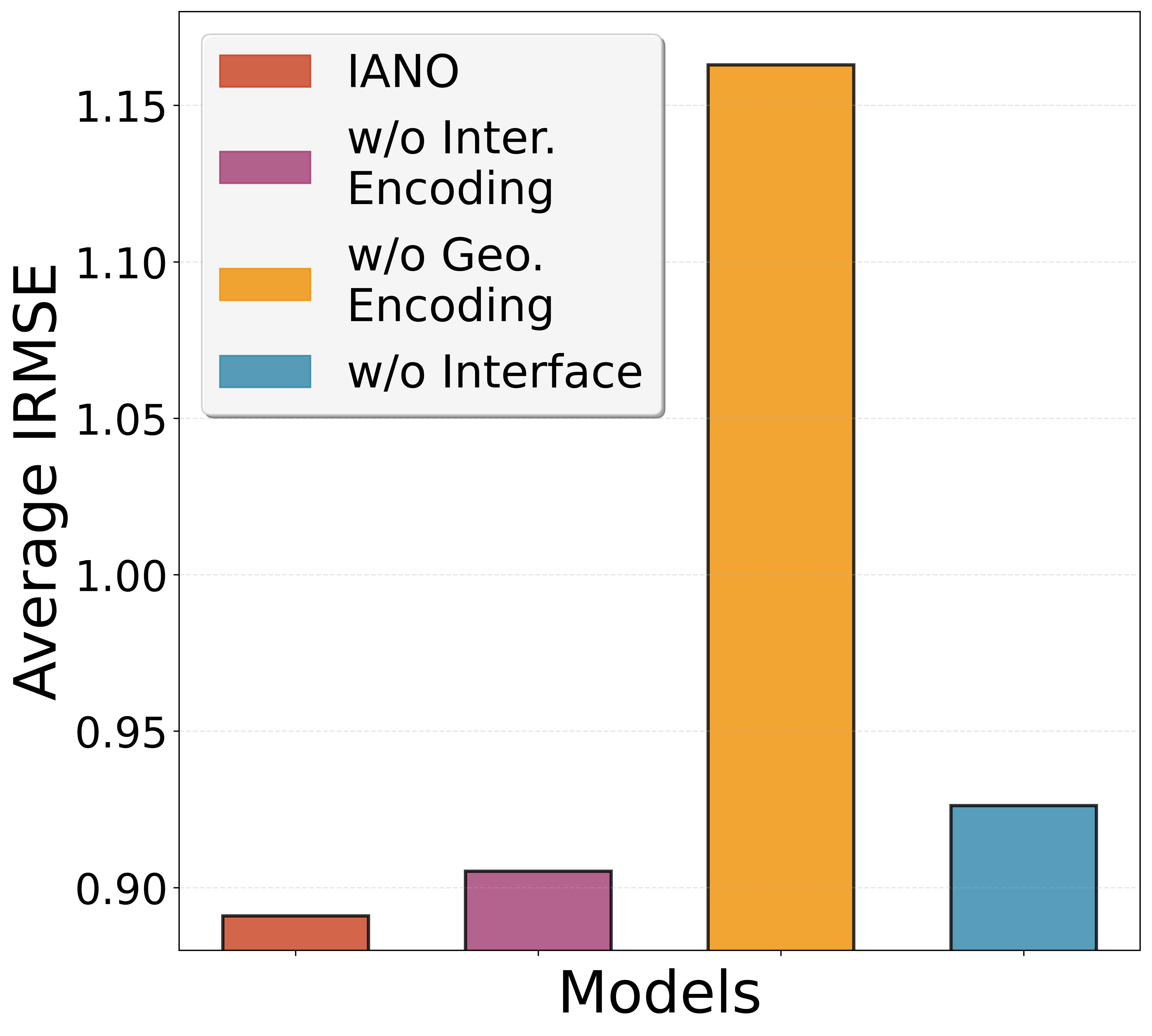}
         \vspace{-0.7cm}
    \caption*{(b)}
  \end{minipage}
         \vspace{-0.25cm}
  \caption{ (a) Noise robustness comparison of various algorithms. (b) Comparison of IANO and its ablated variants.}
   \vspace{-0.54cm}
  \label{fig:robustAndAB}
\end{figure}

\subsection{Robustness under Noisy Data}

This section analyzes the performance of IANO in predicting physical fields under various noise levels, with results summarized in Fig. \ref{fig:robustAndAB} (a). The detailed results are presented in Appendix E. We add Gaussian noise to both the fields and the interface position data at each sampling point. The noise $\epsilon$ is drawn from a normal distribution $\mathcal{N}(0,\sigma^2)$, where the standard deviation $\sigma$ is set to 1\%, 3\%, and 5\% of the original field magnitude, yielding noisy data $\tilde{u} = u + \epsilon$. As depicted in Fig. \ref{fig:robustAndAB}, IANO exhibits superior prediction accuracy across all tested noise levels (1\%, 3\%, and 5\%), achieving optimal results in both RMSE and IRMSE. At a 1\% noise level, IANO recorded an RMSE of 0.364 and an IRMSE of 1.219, outperforming the next best U-Net model. As the noise level increased to 5\%, IANO's performance degradation was relatively minimal, with an RMSE of 0.383 and an IRMSE of 1.223. This sustained superiority in the presence of noisy data underscores IANO's robust capability when faced with noisy data in real-world applications.

\subsection{Ablation Studies}

Our ablation study, briefly shown in Fig. \ref{fig:robustAndAB} (b), evaluates the contribution of each component to IANO. The detailed results are presented in Appendix F. We analyze three ablated variants: `w/o Inter. Encoding' (removing Interface-Aware Multiple Functions Encoding), `w/o Geo. Encoding' (removing Geometry-Aware Positional Encoding), and `w/o Interface` (excluding interface data).
The results show that our full model outperforms all ablated versions across RMSE and IRMSE metrics. 
Notably, (`w/o Geo. Encoding') leads to a significant performance drop, particularly evident in IRMSE for the subcooled and saturated cases. This highlights the critical role of precise geometric context in accurately capturing field values. (`w/o Interface') degrades performance, underscoring the value of directly leveraging interface information for more accurate prediction. (`w/o Inter. Encoding') also causes a performance reduction. 

We also compared the performance of integrating interface data into other methods, such as GNOT and MIONet. We found that their performance showed trivial improvement, or even deteriorated. This is because while MIONet and GNOT can use interface data, their architectures lack the specialized modules necessary to effectively model the intricate coupled relationships between the physical fields and the topological positions of the interfaces, leading to negligible improvement or even degradation. In contrast, our specifically designed modules can effectively handle interface data. A detailed analysis and the results can be found in Appendix F.

\section{Conclusions}
The proposed IANO tackles the data scarcity and spectral bias in neural operators for multiphase flows by explicitly leveraging interface information as complementary variables. It has two main mechanisms: an interface-aware function encoding for multiphysics fields and interface coupling relationship perception, and a geometry-aware positional encoding for spatial fidelity. Experiments show IANO outperforms baselines, achieving $\sim$10\% higher accuracy. Also, IANO proves robust to limited and noisy data. 
The work indicates that leveraging interface data obtained from experimental techniques or numerical simulations as a supplement can mitigate the challenges of low data regimes and spectral bias. This approach could pave an effective

\bibliographystyle{IEEEtran}
\bibliography{aaai2026}

\end{document}